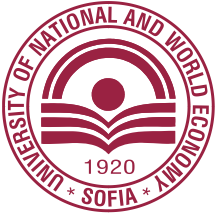

УНИВЕРСИТЕТ ЗА НАЦИОНАЛНО И СВЕТОВНО СТОПАНСТВО
КАТЕДРА „УПРАВЛЕНИЕ"
________________________________________________

BULGARIAN ACADEMIC SIMULATION AND GAMING ASSOCIATION
FRIDAY NIGHT SEMINAR CLUB

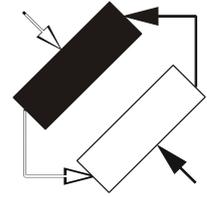

# АВАНГАРДНИ НАУЧНИ ИНСТРУМЕНТИ В УПРАВЛЕНИЕТО

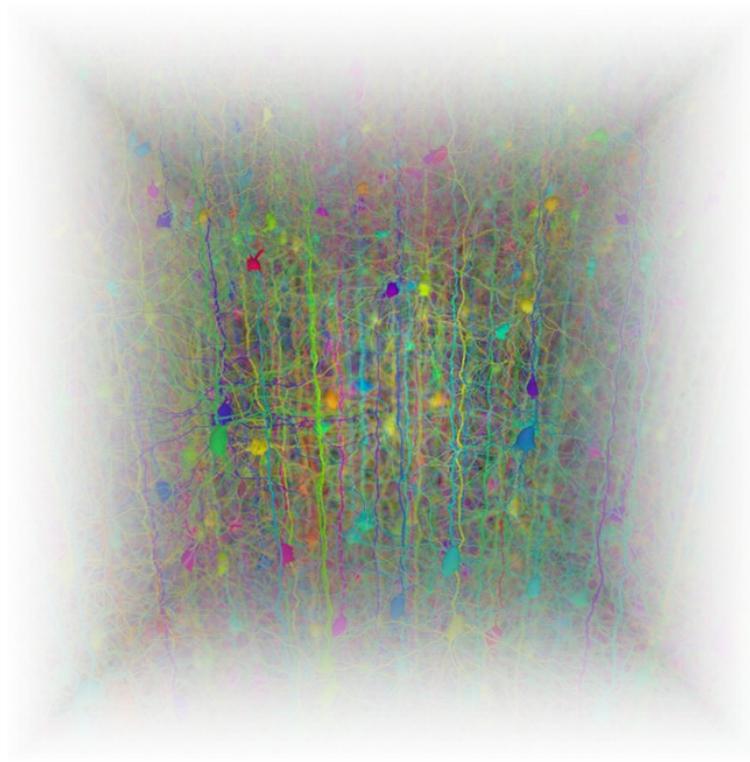

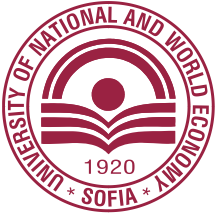

UNIVERSITY OF NATIONAL AND WORLD ECONOMY
DEPARTMENT OF MANAGEMENT
------------------------------------------------
BULGARIAN ACADEMIC SIMULATION AND GAMING ASSOCIATION
FRIDAY NIGHT SEMINAR CLUB

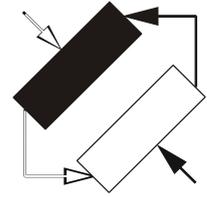

# VANGUARD SCIENTIFIC INSTRUMENTS IN MANAGEMENT

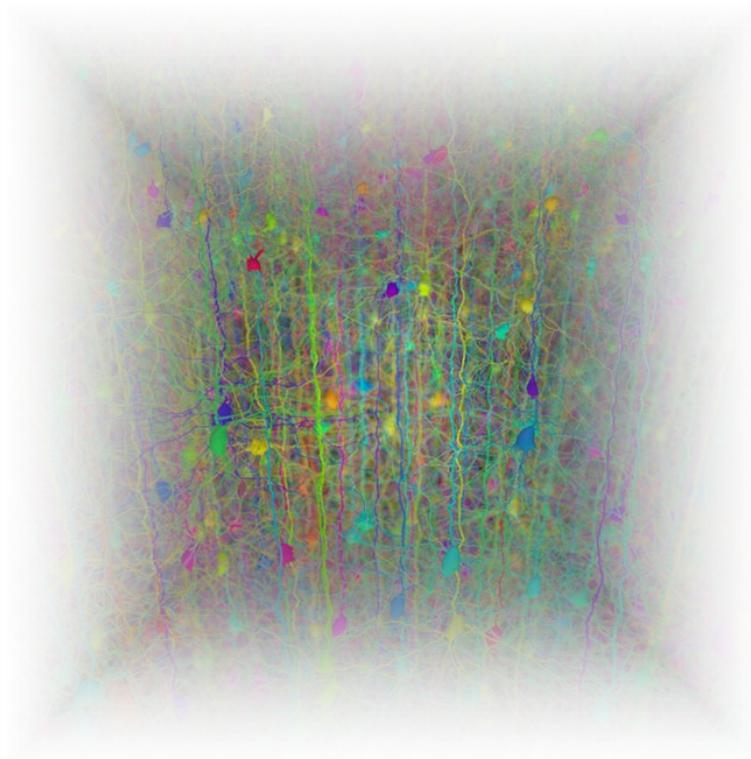

Volume 2(7)/2013                                           ISSN 1314-0582

Sofia, 2013

All authors take full responsibility for the authorship and the originality of the work, as well for mistakes due to their fault. Authors retain all copyrights on their publications.

## АКТУАЛНИ ТЕМИ В СТОПАНСКОТО УПРАВЛЕНИЕ
## CURRENT TOPICS IN BUSINESS MANAGEMENT

**ПРЕДПРИЕМАЧЕСКИЯТ ДУХ В ЕКСПОНЕНЦИАЛНИТЕ ВРЕМЕНА**
**THE ENTREPRENEURIAL SPIRIT IN EXPONENTIAL TIMES**

**ЧОВЕШКИ КАПИТАЛ**
**HUMAN CAPITAL**

## ПУБЛИЧНИ ПОЛИТИКИ И УПРАВЛЕНИЕ
## PUBLIC POLICIES AND GOVERNMENT

# CONTEMPORARY FACETS OF BUSINESS SUCCESSES AMONG LEADING COMPANIES, OPERATING IN BULGARIA

associate professor Kiril Dimitrov, Ph.D.,
"Industrial business" department, UNWE – Sofia[1]

# СТРАТЕГИЧЕСКИ АСПЕКТИ НА УСПЕХИТЕ В БИЗНЕСА СРЕД ВОДЕЩИТЕ КОМПАНИИ, ФУНКЦИОНИРАЩИ В БЪЛГАРИЯ

доц. д-р Кирил Димитров,
катедра "Индустриален бизнес", УНСС – гр. София,

*Abstract*: The current article unveils and analyzes some important factors, influencing diversity in strategic decision-making approaches in local companies. Researcher's attention is oriented to survey important characteristics of the strategic moves, undertaken by leading companies in Bulgaria.

*Резюме*: В тази статия се разкриват и анализират важни фактори, определящи разнообразието в подходите при взимането на стратегически решения в местните компании. Интересът на изследователя е насочен към проучване на значими характеристики на стратегическите ходове, предприемани от водещите компании в България.

*Ключови думи*: корпоративна стратегия, бизнес стратегия, бизнес организации, стратегически мениджмънт, управление на кризи.

## 1. Introduction

The first years of Bulgaria's accession to the European Union and the ongoing effects of the World financial and economic crisis represent the greatest challenges to the managers in the local companies nowadays. While managers' behaviors as expression of complaints and manifestation of humility and stoicism in the turbulent environment are widespread and may be easily monitored in social media, managers' behaviors, associated with business success achievements, still remain to some extent covert. As a rule these behaviors of success are incarnated in the strategic moves, pursued by leading local companies. That is why the last have become of great interest to the researchers and are chosen as a subject of this scientific article.

---

[1] e-mail: kscience@unwe.eu

## 2. Interpretations of strategic management milestones by business practitioners

Dominating views to strategic management in leading organizations in Bulgaria for sure change and to some extent follow the global trends. This phenomenon may be partially explained by: (a) the traditional interest of local society in the world achievements and the strive of applying them at home ground, (b) the acquisition of many local entities after the privatization process by mighty organizations from the developed (or other) countries that to some degree impose the culture of the headquarters on different subsidiaries all over the world. But concerning the field of strategic management the starting point of the transition process to market economy and democratic political system in Bulgaria (i.e. 1990) forms a very low base. The reason for this unfavorable situation lies within the state of executing the tasks of strategic management during the socialist period, as follows:

- *Place of execution*: different units in the respective Bulgarian ministries were responsible for the strategic development of whole national economy sectors, because the organizations, producing products and/or delivering services, were state-owned. The overall company development strategies were transmitted "from the central state administration" and thus at organizational level the production and operation management remained basic responsibility of local managers. Furthermore, the strategic actions of separate cooperatives were directed by the headquarters of the Central cooperative union.

- *Attitudes to cooperation and competition*: cooperation between entities was governed from the level of the state administration and on international level – by the Council of mutual economic assistance (the economic integration structure among the counties from the socialist block). Competition among similar entities in the country did not exist because of the planning, executed at national level that secured definite markets for each entity. Thus, "competition' phenomenon was confined only to these companies that exported their products and/or services outside the iron curtain.

- *Orientation to the utilization of resources*: scarcity of resources was not considered as basic economic paradigm which contributed to domination of low productivity, higher resource consumption (incl. energy, human resources, materials, etc.), technical backwardness and greater environmental pollution by local entities in comparison to leading business organizations from developed countries that were forced to perish or survive in the times of the energy crisis from 1970s, serving as a tipping point for introduction of lasting changes in their strategic leadership principles as pertinent demonstration of value oriented behavior, establishment of the manifestation channels for managers' ethics, bringing out the social responsibility organizational efforts, aspiration after continuous improvement of management methods and mass use of IT, intrepidity to solve complex issues creatively by creation of internally integrated teams, continuous orientation to total quality management and achievement of long-term business related aims. Because of political and economic isolation the local companies postponed the suffering effects of this crisis and could not use it as an overall business improvement opportunity. That is why at the beginning of the transition process their level of competitiveness proved to be lower in comparison to similar organizations from the "capitalist world". This was one of the basic reasons for the deep economic crisis in Bulgaria during the 1990s.

- *The mix between public justifications and latent functions (not spoken of) of the organizational existence*: in many local organizations the respective management teams could not establish a healthy balance in their relations with different constituencies, because strategic management activities were moved out of the entity. So they failed in defining their organization's function in the larger scheme of things and the reasons, justifying its continued existence.

- *The managers as a driving force*: the abandonment of the strategic management responsibilities by the state required urgent enrichment of knowledge and skills inventory at disposal of management teams in local organizations in order to navigate successfully in a market environment. This should be accompanied by formulation and implementation of adequate strategies and development of appropriate organization designs.

Business culture inertia, guarding the set of basic assumptions dominating among managers at that time and serving as a human defense mechanism against experienced fear and anxiety from newness may be indicated as one of the main reasons for impeding fast changes in the entities. Other reasons for demonstrated resistance to change by influential people in organizations may be determined in several realms: (a) the specific individual's psychological capacity of accepting newness in life that traditionally decreases with aging for the majority of working people; (b) lack of knowledge, skills and experience in the sphere of modern organizational change approaches; (c) lack of agreement in society on what to change and how to change it; and (d) persistent defense of the interest of certain stakeholder groups even to the detriment of a target entity (Schein, 2004; Dimitrov, 2012a; Dimitrov, 2012b).

That is why many managers easily found reasons to postpone their strategic decision-making, associated with important and pending organizational issues which caused the emergence of lasting situation when an array of key marker events, reverberating from the business environment, failed in their contribution to the initiation of deliberate processes in many entities with the aim of provoking small and incremental (evolutionary), or deep and fast (radical, revolutionary) desired changes (see figure 1).

That is why country's modernization was hampered to a great extent in the 1990s. Furthermore since the beginning of the 21$^{st}$ century and up to the moment changes in local organizations and society as a whole have not passed through seamlessly due to: (a) the aforementioned cultural inertia among senior managers who as a rule belong to the elder generations with the exception of the IT sector; and (b) the economic effects of the World financial and economic crisis (the first capitalist crisis in Bulgaria's modern history) that contributed to role shift for many contemporary local entrepreneurs who went bankrupt and became again hired laborers, forming the assumption in society that richness has no lifetime guarantee.

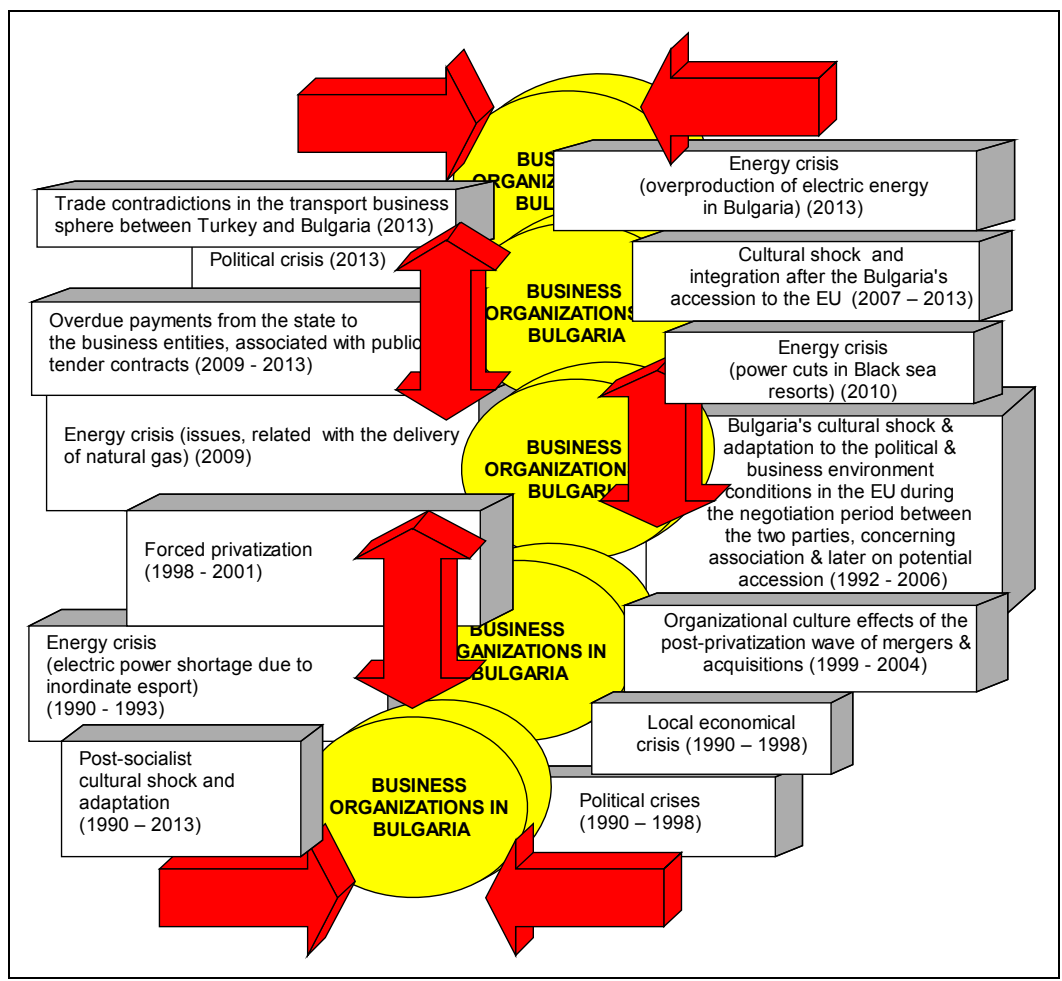

**Figure 1.** The myriad of crises local leading business organization have passed through, since the beginning of the transition

That is why it seems up to par that a large interval outlines diversity in strategic choices, made by management teams due to differentiated assumptions about effectiveness and efficiency of various strategic management paradigms. At least several dimensions of this diversity with direct impact on selection of corporate (business, functional) strategies in leading companies may be traced, as follows:

• *Choosing the right size of the business organization*: It comes to be a tough strategic decision somewhere between the two extremes of a big entity versus and a small one, blazed by alternating impact of characteristics as: (dis)economies of scale, levels of transaction costs, changes in capacity to access outside resources (for instance: capital, bargaining power, etc.), the essence of appropriate organization structure, realized speed of responsiveness to changes, attitudes to innovations, frequency and complexity of arising coordination and communication issues, the ratio in company undertakings between organic growth and growth through mergers and acquisitions, the pursued structure for the portfolio of possessed businesses, differing in economic branches, life cycles and occupied places in the respective value chains, dominating attitude to divestment (including outsourcing and downscoping), and preferred forms of collaboration between the valuable partners, etc. (Helms, 2006; White, 2004; Drucker, 1973). In many local organizations the right-sizing issues are still underestimated. The owners of some privatized entities to some extent inherited their size and most of the time soon after the acquisition initiated projects, oriented to downsizing. As a rule the size of foreign subsidiaries is set by the headquarters.

- *Balancing the interests of different stakeholder groups of the company:* it may become a reality through the sincere and persistent managers' efforts in establishing a desired organization culture by (re)defining appropriate mission, strategy and goals, means of achieving them, ways of measuring the results (including internal control system) and usually chosen corrective actions. Of course it should not be neglected the impact of the capital market on the business organization as a strong lever of imposing strict discipline in thinking and action on different constituencies for the company. In this way the managers occupy a better position in their endeavors to regulate certain behavioral patterns, affecting the way in which strategic process flows in the company, as follows: the level of concentration on achieving strategic targets at the end of a planning period, the habit of scapegoating, formulating less demanding goals, and emphasizing short-term fixes (for instance: underinvestment or misrepresenting performance) (White, 2004; Schein, 2004). The underdeveloped local capital market cannot serve as a control mechanism on managers' activities, oriented to different constituencies and to some extent hampers the managers in their efficient use of evolutionary change mechanisms. The last 23 years in the Bulgaria's economy development have been a period of initial accumulation of capital for the functioning entities, while corporate social responsibility has been perceived as a fad or unwanted movement, imposed by international bodies or the state administration.

- *Selecting the appropriate situations in which it is better for the company to cooperate rather than compete:* current business wheels force organization's management to monitor different types of players on target markets – competitors, suppliers, clients, producers of substitute products from other industries, potential new entrants and (quasi-)government structures. The specific strategic orientation, chosen by the business organization, is always related with bearing of some risk due to occupied competitive position by the respective market leader, other players' perceived and actual strategies, and the environment in which strategic decisions are made. The levels of this risk may be reduced to some extent by organization's active efforts to acquire information about the implemented strategies of target players and timely unveiling of situations when cooperation among players is likely to generate better results for all of them (for instance: weakening price competition, entering a new market, satisfying organization's needs of personnel training, withdrawal of a loan, etc.) in comparison to undertaking of aggressive actions or inefficient free riding by an entity. A cooperation initiative may be led by a single entity, trade association, a group of interested companies or imposed by a (quasi) government body. The real strategic behavior of the contemporary business organization represents an efficient mix of competition and cooperation necessary in most business situations. This is the reason why the term "coopetition" came into being in the management literature (Hill, Jones, 2010; White, 2004). Forming of clusters still is not a wide practice among local companies. Many managers consider it is enough for their entity to participate in just one supply chain or entrepreneurial network or to rely only on one client.

- *Mastering the knack of managing risk:* nevertheless how the term "risk" is defined by the managers in a company – a variance in an important performance indicator or a possibility of an extreme event occurring, its existence should be paid attention to by deliberate design of a risk management strategy that has to be in congruence with the overall organization's development strategy and the information strategy, because resources, efforts and time have to be spent for the acquiring of necessary data, its retrieval, analysis and interpretation. Reducing risk ignorance and concealment is very important because risk and return levels represent the two sides of a coin for appraisal of each potential project for the company. Estimating the risk levels of different situations, contemporary managers wisely use techniques as scenario building and planning to formulate and implement an appropriate mix of generic risk control strategies, i.e. (a) avoidance strategy – always used after detailed analysis that brought to rejection or termination of a project (b) mitigation strategy, applying negotiation, flexibility of any arrangements or diversification as methods for risk reduction, and (c) risk management strategy that means sharing of a fixed risk by a commercial arrangement, i.e. insurance or hedging, inclusion in a strategic alliance, or securing governmental support (Dobson, Starkey, Richards, 2004; White, 2004). Risk management culture among the majority of executives is still at an unsatisfactory level and frequently there can be heard voices, lobbying the state administration to help in development of their industries.

- *Developing the adroitness in carrying out of new market entries, especially at global level:* the sharp clashes, caused by the impacts of national and regional cultural differences tempt business organizations to pursue both higher return and lower risk, applying an array of specific and repeated strategies for starting their participation in global business, varying from exporting, licensing or franchising, foreign direct investment, joint ventures, acquisitions or mergers and green-field projects. All these differ in their fitting to the specific milieu in the respective company and the respective business environments the entity intends to operate in, i.e. degree of resource commitment (availability of resources) and levels of (and managers' attitudes to) return and risk (risk-averse or risk-oriented), diversification necessities and abilities, level of transaction costs, associated with switching over different entry modes and the degree of home country bias, level of defense for intellectual property rights of mobile, created assets (Schermerhorn, 2012; White, 2004; Ivancevich, Lorenzi, Skinner, Crosby, 1994). On one hand, during the transition period Bulgaria was forced to bankrupt most of its internationalized organizations, while saving some (parts) of them through privatization. On the other hand, some newly created private companies tried to internationalize their activities by entering new markets, but the effects of the World economic and financial crisis froze these endeavors, except the cases when an entity belonged to a counter-cyclical industry.

- *The attributes, directing the specific flow of strategic management process in the business organization*: some of the nuances in performing certain analyses and/or choosing concrete strategies may be explained by the complexity of the respective organization and the position in it of a definite business unit (see figure 2) which gives the managers the right to pose specific questions (How do we make money? Which industries should we be in? How should we compete?). Furthermore, managers in different organizations may hold different assumptions of how to realize strategic management process, based on their adoption of certain management school doctrine. For example a sharp difference is observed at corporate level where dominating concepts of roles and interactions among senior managers and boards' directors in big companies direct how strategies are designed and implemented on all levels in the entity (corporate, business, functional) and how an organization's partnerships are carried out (network level). On one hand the strategy pyramid approach strictly adheres to the top-down direction in its formulation process, as follows: vision, mission, goals, strategies, tactics and action plans. In this way the entity may be effective only in stable environment, predominantly relying on its existing competencies. On the other hand strategy formulation process may be stretched simultaneously in two directions (top-down and bottom-up). In this way the managers may detect emerging opportunities for the company as a whole and some of its business units in unstable environments or entering in a new one. Thus, the organization has a chance to learn new competences while reformulating its strategic intents on its challengeable road of development (Schermerhorn, 2011; Kotelnikov, 2006; Mintzberg, Ahlstrand, Lampel, 1998). Except the ICT industry, most of local managers have long lived during the socialist times and to some extent still hold basic assumptions of top-down direction in strategy formulation as the only one possible way of getting work done and obligatory implementation of strong and heavy organizational designs as means to achieve entity's targets.

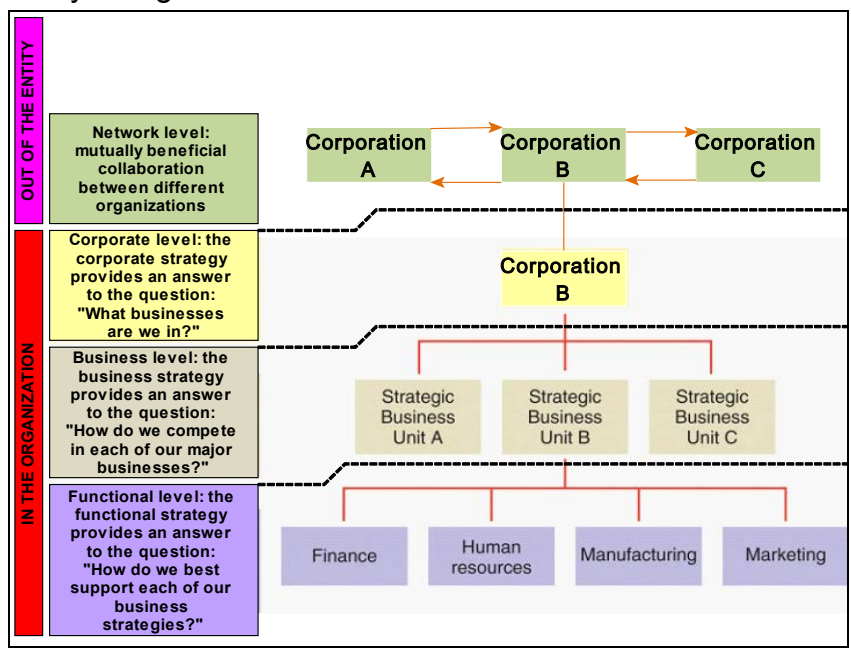

Source: adapted from Schermerhorn, J., Introduction to management, 11[th] edition, John Wiley & Sons (Asia), INC., 2011., p. 210

**Figure 2.** The four levels of strategy (in and outside the organization)

- *The characteristics of the shared meaning for the term "strategy":* managers' definition of the contemporary business environment, based on observed and/or perceived levels of (in)predictability, (in)stability, (un)certainty, the speed of wealth accumulation and loss influences preferred attributes of chosen strategies from the company, i.e. (a) along the continuum "simplicity – complexity': It describes in essence and numbers the strategic processes, pushing the company in pursuing the most promising opportunities, and the established rules, driving these processes, based on managers' inclinations to rush at insecure situations where opportunities are great and diverse. (b) by the speed of arising changes in the business environment: the higher the aforementioned speed, the better for the managers to design the overall company development strategy as unique deployment of separate business units by appropriate opportunities, choosing the right size and focus for each unit. That is why the managers are forced to view at their teams of employees as mixes of knowledge, skills and capabilities. In this situation the managers first select the right people and then assign the roles to the appropriate ones. (c) by duration of strategy: the fast-changing markets constrain managers to search continuously unique strategic changes for products, services, technologies, locations, etc. in order to timely detect, utilize and abandon certain opportunities. (d) managers' ability to pursue simultaneously more than one strategic approach (for instance: adhering to the core competences and creative destruction); (e) managers' attitude to undertaking changes (by magnitude: small, incremental; medium, big, radical, deep; and by frequency: usual, from time to time, rare) (Eisenhardt, 2002; Campbell, Stonehouse, Houston, 2002). Many managers in Bulgaria still consider all activities that are not directly related to production and sales short-term revenue, a total loss of time and underestimate their long-term contribution to company's profit margins, as the case of strategic management looks like.

- *Dominating understandings about the essence of the work that senior managers (especially the board members) should perform:* what is meant here is the achieved performance levels by the board members of the company in specific tasks: (a) ensuring the strategic planning process flow in the entity, making and applying in their work deliberate choices, monitoring the implementation of current strategic initiatives; (b) monitoring the company's succession planning system; (c) verifying the right functioning of company's information, control and audit system; (d) guaranteeing company's compliance with legal and ethical standards, imposed by laws and firm's mission; (e) avoiding and managing crises; (f) improving the board's functioning (Sonnenfeld, 2002; Conger, Finegold, Lawer III, 1998). Many organizations in Bulgaria have a dominating blame culture, characterized by smoldering conflicts between hired laborers and management, because of unsatisfactory workplace conditions, and great differences between payment levels of the two groups.

- *The postulates of contemporary strategic paradigm for sustainable business success achievement - comprehension, attitudes and degree of use in routine work by managers which may be expressed by:* (a) finding the right mix among organization's activities, oriented to adaptation and change in response to the external environment and those, focused on the dynamics of the internal integration of systems, structures, and processes in the entity; (b) establishing an appropriate prognostication/ planning horizon for undertaken firm's initiatives to balance successfully between the pursuing of short-term financial goals and a long-term vision; (c) deliberate implementation of organization structures with appropriate characteristics along the continuum, formed by high flexibility and great opportunities for timely decision-making against heavy bureaucracy and realization of retarded reactions to environmental phenomena; (d) forming the right attitude to status-quo along the continuum of contentedness against dissatisfaction in order to shift the organization towards embracement of better opportunities or minimizing the risks, associated with entity's strategic development (Denison, 2010; Daniell, 2004). The local political and economic crisis from the 1990s established a short-term horizon in managers' endeavors that aggravates company market performance.

- *Mastery and use in everyday work of modern strategic management terms by the executives*: on one hand, the failures and poor performance of entities in the undertaken projects to a great extent are due to the extensive use of old-fashioned strategic approaches, models, moves and professional language, related with them. On the other hand, deliberate apprehension, strong adoption and wide use in routine work of new strategic approaches or enriched understandings of strategy essence and contents seems to be strongly related with study of "a new language". Dynamics of contemporary business environment is associated with terms as "industry profit pool", "business process portfolio" or application of creative job titles as "chairman and chief software architect", "chairman and chief mentor". Modern terms as globalization, complexity, dynamism, turbulence, acceleration, rationalization, obsolescence and reinvention, connectivity, convergence, ephemeralization (moving to the virtual), consolidation, etc. are frequently used by senior managers in leading organizations from the developed countries. The unlearning of old terms by senior managers comes to be the other side of the coin. For example in some companies the executives avoid to use the term "profit" and replace it with "intermediate contribution" or "notional result". But language only is not enough, because its use has to be tightly bound to certain metrics in the established performance measurement systems (Katsioloudes, 2006; Daniell, 2004).

Most of the aforementioned determinants of strategic management cannot be explored directly because of company security issues, but their effects on company performance may be traced to some extent by content analysis of publicly shared information about undertaken strategies by leading business organizations in Bulgaria.

### 3. Methodology of the survey

The dominating "company security" paradigm in the minds of business managers - for sure based on stable grounds, hampers researchers in their quest of penetrating the sphere of strategic management practices among the local companies. That is why this survey chooses an indirect approach in supplying needed data for analysis of strategic moves, undertaken by the successful business organizations, operating in Bulgaria, i.e. acquiring of respective market information, published in different annual firm lists of leading local economic periodicals during the calendar year of 2012. It has to be mentioned that this information actually presents company data from the previous fiscal year, i.e. 2011. In this way the survey may reveal important characteristics of company behavior in the current turbulent times, marked by the numerous influences of the World financial and economic crisis on local entities and their constituencies (2013a; 2013b; 2012c; 2012d; 2012e; 2012f). Presented in writing by economic journalists interviews with managers of leading business organizations and economic analyzes of the market performance of key players in target industries included in the enclosures, showing succeeding companies lists, may be defined as primary data sources in this survey. Different lists are intended with different target groups of companies, i.e. the big ones, the small and medium sized ones, the entities in a given industrial sector which provides a greater diversity of objects to this survey. In details the needed data is acquired by several techniques: (a) direct citing senior managers' words from their interviews in the enclosures of the respective periodicals; (b) data, collected and retrieved from journalists' articles, dedicated to target sector analysis, included in these enclosures; (c) researcher's theoretical inferences based on company key historical events, made public by senior managers in their interviews or unofficial talks with the journalists.

With the aim of securing a larger number of surveyed companies, the author does not impose any restrictions by branches of industry in which the firms may operate or by entity's size, measured by the indicator of average number of personnel. The main research question is defined, as follows: "How can be described the main facets of strategic management, executed in the companies, operating in Bulgaria?". Finding a satisfactory answer requires decomposing it to investigative questions, as follows: (a) "What are the contemporary strategic moves of the succeeding business organizations in Bulgaria?". As a prerequisite to acquire richer information through posing this question, seven reliable sources of strategic management are used as a theoretical background (Porter, 2008; Ritson, 2008; Dobson, Starkey, Richards, 2004; Grundy, 2003; Campbell, Stonehouse, Houston, 2002; Hristov, 2009; Paunov, 1995). (b) What is the number of the mentioned strategic moves by interviewed managers as a part of the basic bundle of overall company's development strategy? (c) Do the managers mention any specific facets, related with the basic bundle of overall company's development strategy and how many? (d) Is the company possessed by another mightier organization? And what is the origin of the mother company? (e) Do managers of leading organizations use modern strategic management related terms as a part of their professional language? (f) What additional company related data can the researcher collect? (i.e., the branch of industry the respective organization belongs to, the survey where the company information stems from, the name of the company, its size, and year of the respective survey).

The scientific approach to answering the basic research question and the related with it investigative questions requires a formulation of a strict definition for the term "a succeeding company", depicting the compulsory shades of meaning, as follows:

- A succeeding business organization may be an entity that had such an increase in its profit for the last period under review (the fiscal year of 2011) that it is enlisted for the first time in lists of succeeding companies or it has occupied a better position in comparison the previous period under review.

- A succeeding business organization may be an entity that underwent a certain sales decrease during the previous fiscal period, but the observed shrinkage was the smallest in comparison to its competitors or the shrinkage might be larger, but the organization fulfills the criteria for inclusion in the respective list.

This is how needed information of overall company strategy and its related facets is collected. Of course this information is officially proclaimed by the senior managers in these organizations and this fact may give way to deliberate exaggeration of certain facts and passing over in silence of others, considered with a potential impact of ruining the public image of an entity. Consequently, managers as all the people are inclined to overstate their successes which approach is often considered as an element of the public relations strategy, followed by the respective business organization. On one hand the cultural perspective permits adopting the acquired information to some extent as a saga or a legend. On the other hand the number of the companies, included in succeeding entities annual lists, which managers refused to provide official data for their activities that brought to achievement of such results, is great. The reasons for demonstration of similar behaviors may be different, as follows: (a) inability to use efficiently the strategic positions, conquered by the business organization, as a maneuvering ground for achievement of greater successes in the near future; (b) fear of competitors' retaliation activities; (c) dominating leadership crisis in a transition society; (d) the maintenance of underdeveloped bundle of the overall company strategy, lacking its public relations functional strategy; (e) fear for managers' personal life and the physical survival of their family members, etc. That is why several limitations, associated with this survey, are determined, as follows:

- The firm documents, related with the implemented strategies, are out of reach for the researcher because as a rule all these are defined as "sensitive information" for the respective company. That is why the researcher is forced to base his to some degree subjective conclusions on the data, revealed in presented interviews and industry analyses – especially on the expressed views in them by managers and economic journalists. Of course the opinions of the aforementioned constituencies should be appraised by the researcher through the lens of his deeper knowledge about the impact factors in the local business environment.

- Unofficial information, related with the undertaken strategic moves by the succeeding organizations in Bulgaria, is not available to the researcher and the survey relies only to the aforementioned data sources.

- The population of the surveyed companies constitutes only organizations which managers are not afraid of sharing their success histories with the local community in search of its members' respect, admiration and support. In this way the managers deliberately strive to realize their engagement in the corporate responsibility sphere.

In this way a group of 72 leading business organizations to be surveyed is formed. They belong to different industries of the economy (for example metallurgy, foodstuff industry, construction, machine-building, high-tech sector, chemical industry, etc.).

### 4. Results of the survey

A great deal of respondents from the surveyed business organizations marked out more than one strategy that could be attributed to wide-known theoretical classifications. Thus, the most preferred strategic moves by the managers of the leading business organizations in Bulgaria are "Production/ service related investments" (16.9%), "Export" (12.7%), "High quality standards" (8.5%), Employee training (6.3%), Innovation strategy (5.3%), and Corporate social responsibility (5.3%) (see table 1, figure 3). Thus, a specific combination of the aforementioned strategies and others for each entity is formed to frame the implemented unique strategic bundles, representing the respective overall company development strategy. There is no further information about the extent to which each of the elements in the strategic bundle is pursued by the respective management team.

**Table 1.** Frequencies of strategic moves, undertaken by the companies during 2011 ($str_moves Frequencies)

|  |  | Responses N | Responses Percent | Percent of Cases |
|---|---|---|---|---|
| Strategic moves of the companies (a) | Export | 24 | 12.7% | 33.3% |
|  | Cost optimization/ leadership | 2 | 1.1% | 2.8% |
|  | Innovation strategy | 10 | 5.3% | 13.9% |
|  | Production/ service related investments | 32 | 16.9% | 44.4% |
|  | Marketing related investments | 3 | 1.6% | 4.2% |
|  | Related diversification | 8 | 4.2% | 11.1% |
|  | Unrelated diversification | 7 | 3.7% | 9.7% |
|  | Fulfillment of ecological requirements | 8 | 4.2% | 11.1% |
|  | Position defense | 5 | 2.6% | 6.9% |
|  | Organizational restructuring | 1 | .5% | 1.4% |
|  | Market development | 4 | 2.1% | 5.6% |
|  | Employee training | 12 | 6.3% | 16.7% |
|  | EU funding/ loans | 8 | 4.2% | 11.1% |
|  | High quality standards | 16 | 8.5% | 22.2% |
|  | Reasonable pricing | 4 | 2.1% | 5.6% |
|  | Increase in productivity | 4 | 2.1% | 5.6% |
|  | Sustainable development strategy | 4 | 2.1% | 5.6% |
|  | Corporate social responsibility | 10 | 5.3% | 13.9% |
|  | Differentiation strategy | 1 | .5% | 1.4% |
|  | Product development | 3 | 1.6% | 4.2% |
|  | Outsourcing strategy | 8 | 4.2% | 11.1% |
|  | Hiring new personnel | 6 | 3.2% | 8.3% |
|  | Focus strategy | 7 | 3.7% | 9.7% |
|  | Franchise strategy | 2 | 1.1% | 2.8% |
| Total |  | 189 | 100.0% | 262.5% |

a Dichotomy group tabulated at value 1.

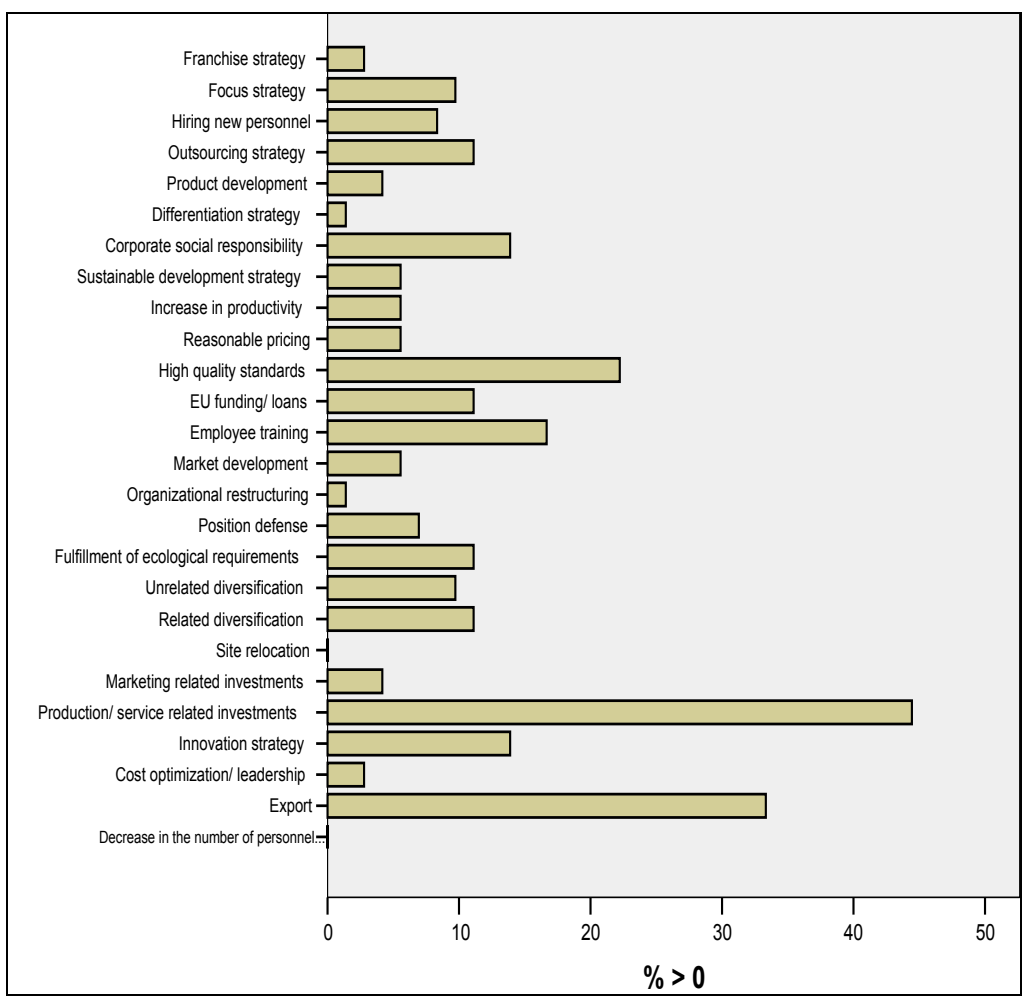

**Figure 3.** Frequencies of strategic moves, undertaken by the companies during 2011 (in percent).

Half of the business organizations proved to be a possession of mightier companies. In this way the relations between the headquarters and the respective business units come to surface, implying that the strategies implemented in the local subsidiary for sure are in congruence with those, followed at corporate level. The last may be used as an explanation to demonstrated specific market behavior by these subsidiaries whose managers have to act as members of a greater network. Most of the members from the last group of entities have mother companies with a foreign origin (75%) (see table 2 and figure 4).

**Table 2.** Surveyed business organizations, classified by the origin of the mother company.

|  |  | Frequency | Percent | Valid Percent | Cumulative Percent |
|---|---|---|---|---|---|
| Valid | Bulgarian | 9 | 12.5 | 25.0 | 25.0 |
|  | Foreign | 27 | 37.5 | 75.0 | 100.0 |
|  | Total | 36 | 50.0 | 100.0 |  |
| Missing | System | 36 | 50.0 |  |  |
| Total |  | 72 | 100.0 |  |  |

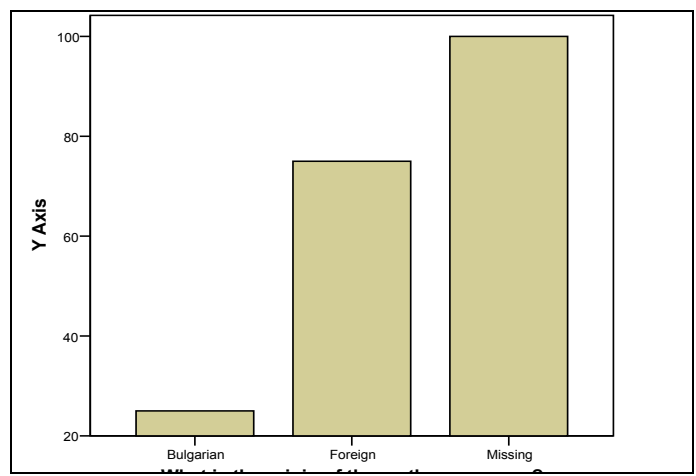

**Figure 4.** Surveyed business organizations, classified by the origin of the mother company.

**Table 3**. The strategic moves of surveyed business organizations, classified by the availability of a mother company

|  |  |  | Is the company possessed by another mightier organization? |  | Total |
|---|---|---|---|---|---|
|  |  |  | No | Yes |  |
| Strategic moves multiresponse | Export | Count | 13 | 11 | 24 |
|  | Cost optimization/l | Count | 1 | 1 | 2 |
|  | Innovation strategy | Count | 5 | 5 | 10 |
|  | Production/ service | Count | 16 | 16 | 32 |
|  | Marketing related in | Count | 1 | 2 | 3 |
|  | Related diversificat | Count | 7 | 1 | 8 |
|  | Unrelated diversific | Count | 5 | 2 | 7 |
|  | Fulfillment of ecolo | Count | 5 | 3 | 8 |
|  | Position defense | Count | 1 | 4 | 5 |
|  | Organizational restr | Count | 0 | 1 | 1 |
|  | Market development | Count | 1 | 3 | 4 |
|  | Employee training | Count | 5 | 7 | 12 |
|  | EU funding/ loans | Count | 7 | 1 | 8 |
|  | High quality standar | Count | 5 | 11 | 16 |
|  | Reasonable pricing | Count | 3 | 1 | 4 |
|  | Increase in producti | Count | 2 | 2 | 4 |
|  | Sustainable developm | Count | 1 | 3 | 4 |
|  | Corporate social res | Count | 3 | 7 | 10 |
|  | Differentiation stra | Count | 0 | 1 | 1 |
|  | Product development | Count | 2 | 1 | 3 |
|  | Outsourcing strategy | Count | 3 | 5 | 8 |
|  | Hiring new personnel | Count | 4 | 2 | 6 |
|  | Focus strategy | Count | 5 | 2 | 7 |
|  | Franchise strategy | Count | 1 | 1 | 2 |
| Total |  | Count | 36 | 36 | 72 |

Percentages and totals are based on respondents.

a. Dichotomy group tabulated at value 1.

The difference in strategic choices made by the management teams becomes evident when preferred strategic moves are cross-tabulated with the availability of a mother company for the respective surveyed entity. The results show that the managers of leading business organizations that are not subordinated to larger corporations do not apply "Differentiation strategy" and "Organizational restructuring". In this group of entities there is wider implementation of "(un)related diversification", "fulfillment of ecological requirements", "EU funding/loans", "Reasonable pricing", "Hiring new personnel" and "Focus strategy" in comparison to the subsidiaries. The business units, dependent on their headquarters general strategies and policies, reveal certain specificity in the sphere strategic management execution, expressed by wider use of "Marketing related investments", "Position defense", "Market development", "Employee training", "High quality standards", "Sustainable development strategy", "Corporate social responsibility" and "Outsourcing strategy" (see table 3 and figure 5)

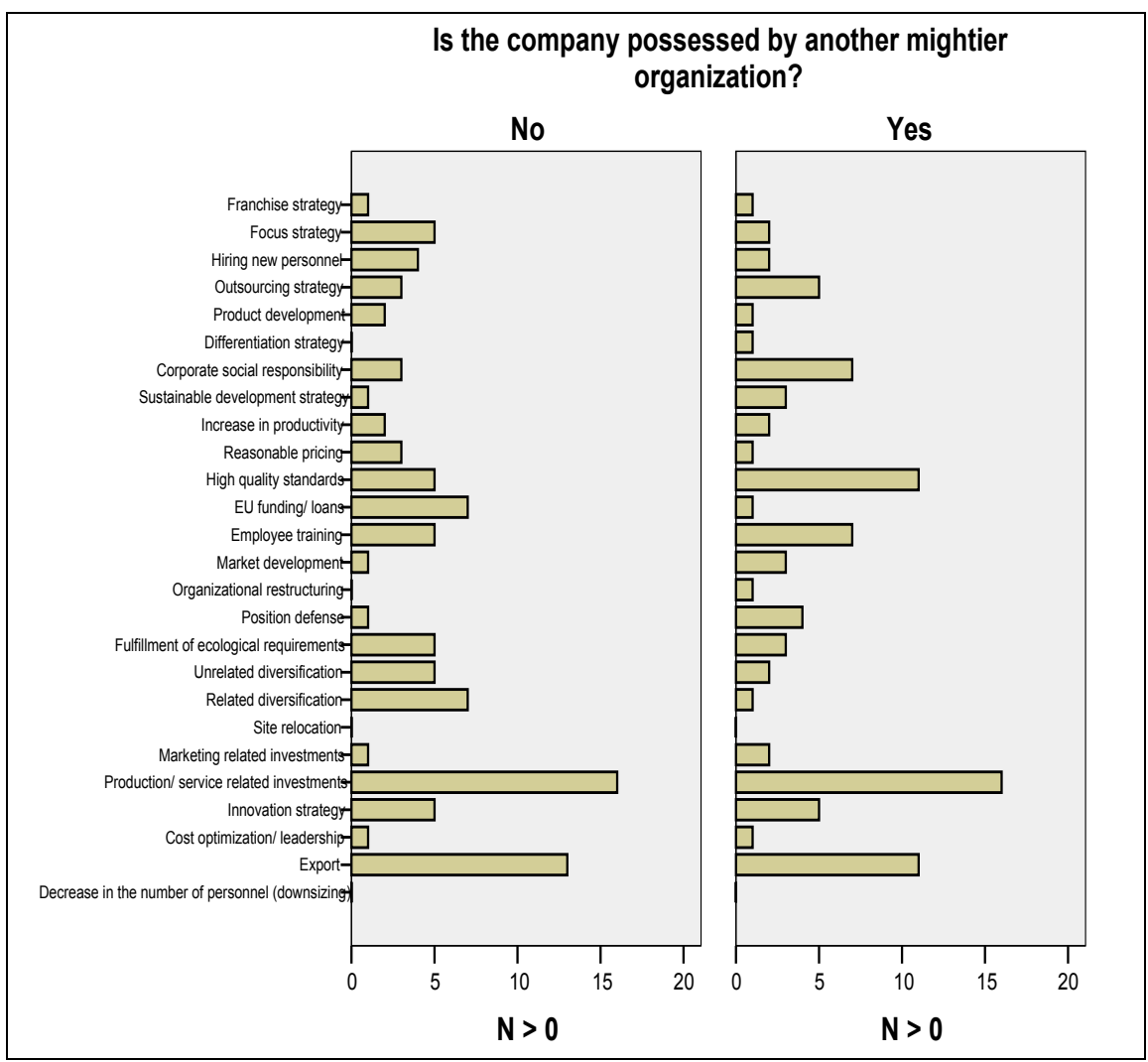

**Figure 5.** The strategic moves of surveyed business organizations, classified by the availability of a mother company.

The subsidiaries with local mother companies claim to pursue narrower array of business strategies in comparison to entities, owned by foreign corporations. The managers in the entities, possessed by local big organizations claim that they do not implement "Franchise strategy", "Hiring new personnel", "Product development", "Differentiation strategy", "Corporate social responsibility", "Increase in productivity", "Reasonable pricing", "Employee training", "Market development", "Position defense", "Fulfillment of ecological requirements", "Site relocation", "Marketing related investments", "Innovation strategy" and "Decrease in the number of personnel". The headquartered abroad entities rely to a greater extent to "Production/ service related investments", "Export", "High quality standards", "Corporate social responsibility", "Employee training", "Innovation strategy" and others (see table 4 and figure 6).

**Table 4.** The strategic moves of surveyed business organizations, classified by the origin of mother companies.

|  |  |  | What is the origin of the mother company? |  | Total |
|---|---|---|---|---|---|
|  |  |  | Bulgarian | Foreign |  |
| Strategic moves by headquarter's origin | Export | Count | 1 | 10 | 11 |
|  | Cost optimization/l | Count | 1 | 0 | 1 |
|  | Innovation strategy | Count | 0 | 5 | 5 |
|  | Production/ service | Count | 3 | 13 | 16 |
|  | Marketing related in | Count | 0 | 2 | 2 |
|  | Related diversificat | Count | 1 | 0 | 1 |
|  | Unrelated diversific | Count | 1 | 1 | 2 |
|  | Fulfillment of ecolo | Count | 0 | 3 | 3 |
|  | Position defense | Count | 0 | 4 | 4 |
|  | Organizational restr | Count | 1 | 0 | 1 |
|  | Market development | Count | 0 | 3 | 3 |
|  | Employee training | Count | 0 | 7 | 7 |
|  | EU funding/ loans | Count | 1 | 0 | 1 |
|  | High quality standar | Count | 3 | 8 | 11 |
|  | Reasonable pricing | Count | 0 | 1 | 1 |
|  | Increase in producti | Count | 0 | 2 | 2 |
|  | Sustainable developm | Count | 1 | 2 | 3 |
|  | Corporate social res | Count | 0 | 7 | 7 |
|  | Differentiation stra | Count | 0 | 1 | 1 |
|  | Product development | Count | 0 | 1 | 1 |
|  | Outsourcing strategy | Count | 2 | 3 | 5 |
|  | Hiring new personnel | Count | 0 | 2 | 2 |
|  | Focus strategy | Count | 2 | 0 | 2 |
|  | Franchise strategy | Count | 0 | 1 | 1 |
| Total |  | Count | 9 | 27 | 36 |

Percentages and totals are based on respondents.
  a. Dichotomy group tabulated at value 1.

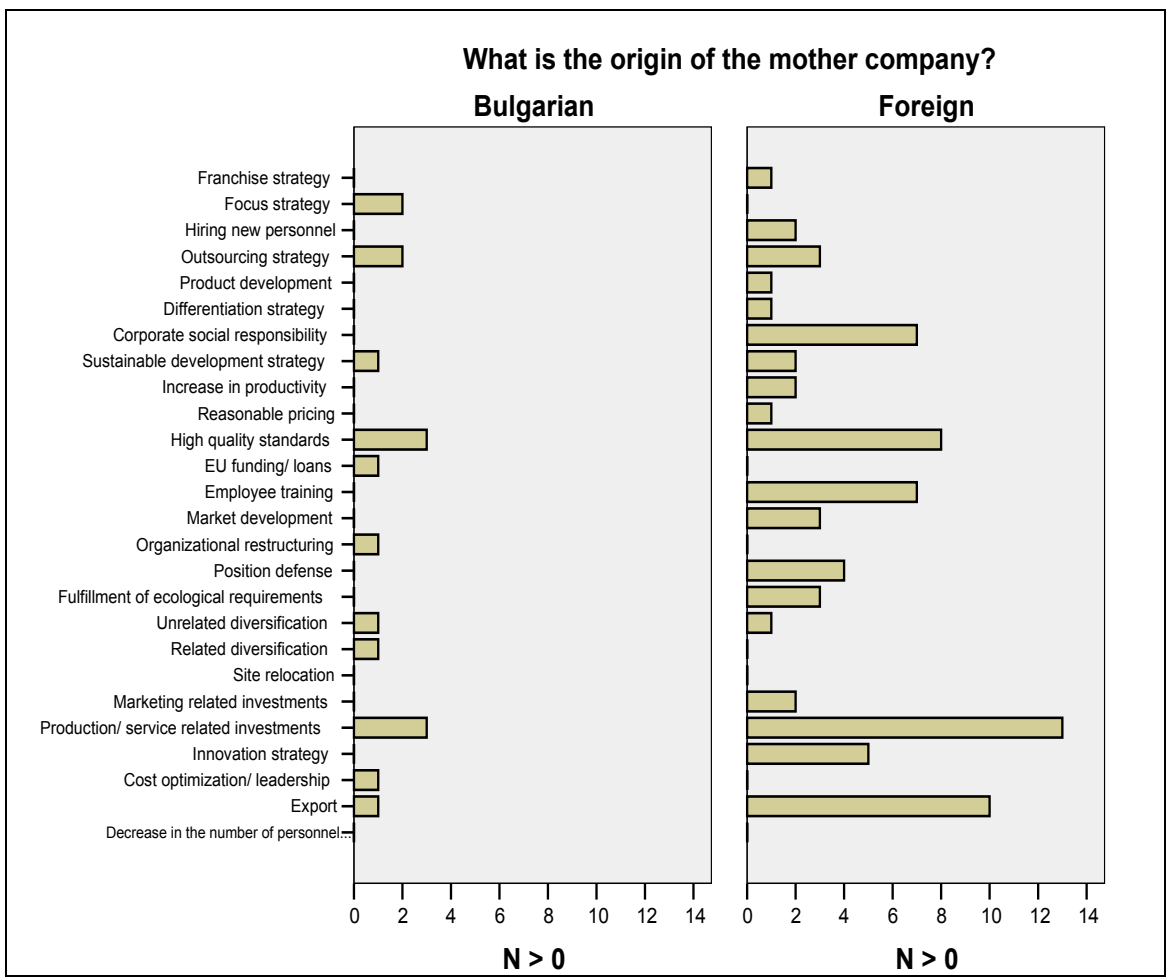

**Figure 6.** The strategic moves of surveyed business organizations, classified by the origin of mother companies.

**Table 5.** Company size

|       |             | Frequency | Percent | Valid Percent | Cumulative Percent |
|-------|-------------|-----------|---------|---------------|--------------------|
| Valid | Big company | 28        | 38.9    | 38.9          | 38.9               |
|       | SME         | 44        | 61.1    | 61.1          | 100.0              |
|       | Total       | 72        | 100.0   | 100.0         |                    |

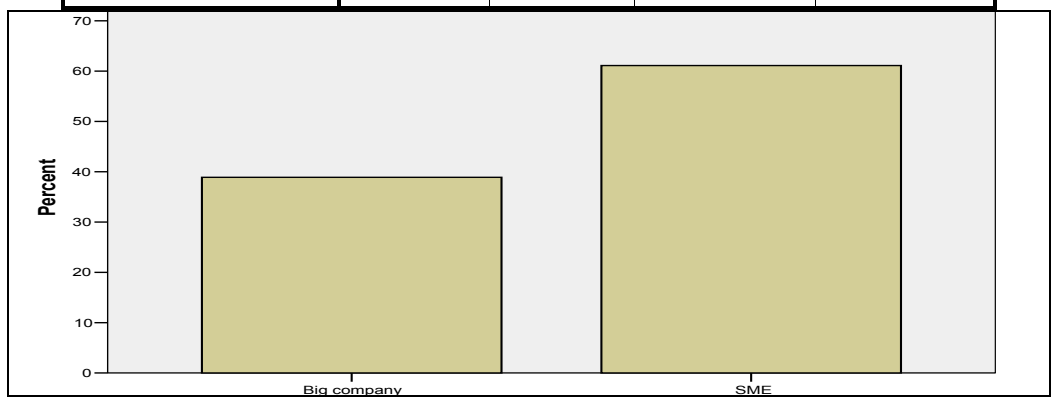

**Figure 7.** Company size data

The small and medium sized business organizations dominate among the group of surveyed leading local entities, because of the structure in the enclosures, presenting the annual lists of leading local companies that is chosen by the respective editors (see table 5 and figure 7).

**Table 6.** The strategic moves of surveyed business organizations, classified by their size

| | | | Company size | | |
|---|---|---|---|---|---|
| | | | Big company | SME | Total |
| Strategic moves by headquarter's origin | Export | Count | 9 | 15 | 24 |
| | Cost optimization/l | Count | 0 | 2 | 2 |
| | Innovation strategy | Count | 7 | 3 | 10 |
| | Production/service | Count | 18 | 14 | 32 |
| | Marketing related in | Count | 2 | 1 | 3 |
| | Related diversificat | Count | 2 | 6 | 8 |
| | Unrelated diversific | Count | 3 | 4 | 7 |
| | Fulfillment of ecolo | Count | 7 | 1 | 8 |
| | Position defense | Count | 4 | 1 | 5 |
| | Organizational restr | Count | 0 | 1 | 1 |
| | Market development | Count | 4 | 0 | 4 |
| | Employee training | Count | 9 | 3 | 12 |
| | EU funding/loans | Count | 3 | 5 | 8 |
| | High quality standar | Count | 12 | 4 | 16 |
| | Reasonable pricing | Count | 1 | 3 | 4 |
| | Increase in producti | Count | 3 | 1 | 4 |
| | Sustainable developm | Count | 3 | 1 | 4 |
| | Corporate social res | Count | 7 | 3 | 10 |
| | Differentiation stra | Count | 1 | 0 | 1 |
| | Product development | Count | 2 | 1 | 3 |
| | Outsourcing strategy | Count | 3 | 5 | 8 |
| | Hiring new personnel | Count | 2 | 4 | 6 |
| | Focus strategy | Count | 1 | 6 | 7 |
| | Franchise strategy | Count | 1 | 1 | 2 |
| Total | | Count | 28 | 44 | 72 |

Percentages and totals are based on respondents.
   a. Dichotomy group tabulated at value 1.

The big business organizations rely to a greater extent to strategic moves as "Corporate social responsibility", "High quality standards", "Employee training", "Fulfillment of ecological requirements" and "Innovation strategy" in comparison to the small and medium-sized ones. Both small and medium-sized entities and big companies emphasize the implementation of "Production/ service related investments" and "Export", and restrain from the use of "Site relocation" and "Decrease in the number of personnel" (see table 6 and figure 8).

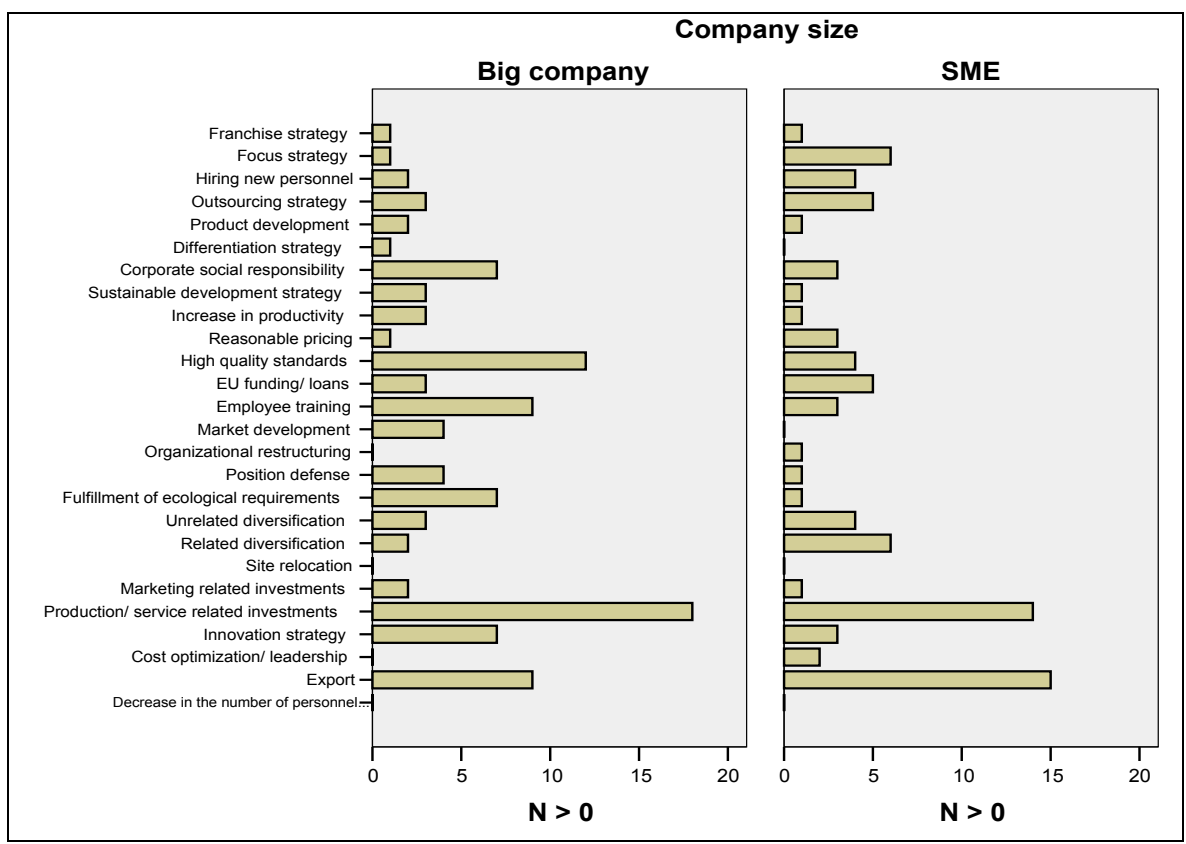

**Figure 8.** The strategic moves of surveyed business organizations, classified by their size

The interval, formed by the numbers of mentioned strategic moves by interviewed managers as a part of the basic bundle of overall company's development strategy is shown in table 7 and figure 9.

**Table 7**. The Number of mentioned strategic moves by interviewed managers as a part of the basic bundle of overall company's development strategy

|       |                      | Frequency | Percent | Valid Percent | Cumulative Percent |
|-------|----------------------|-----------|---------|---------------|--------------------|
| Valid | one strategic move   | 23        | 31.9    | 31.9          | 31.9               |
|       | two strategic moves  | 21        | 29.2    | 29.2          | 61.1               |
|       | three strategic moves| 9         | 12.5    | 12.5          | 73.6               |
|       | four strategic moves | 5         | 6.9     | 6.9           | 80.6               |
|       | five strategic moves | 10        | 13.9    | 13.9          | 94.4               |
|       | six strategic moves  | 3         | 4.2     | 4.2           | 98.6               |
|       | seven strategic moves| 1         | 1.4     | 1.4           | 100.0              |
|       | Total                | 72        | 100.0   | 100.0         |                    |

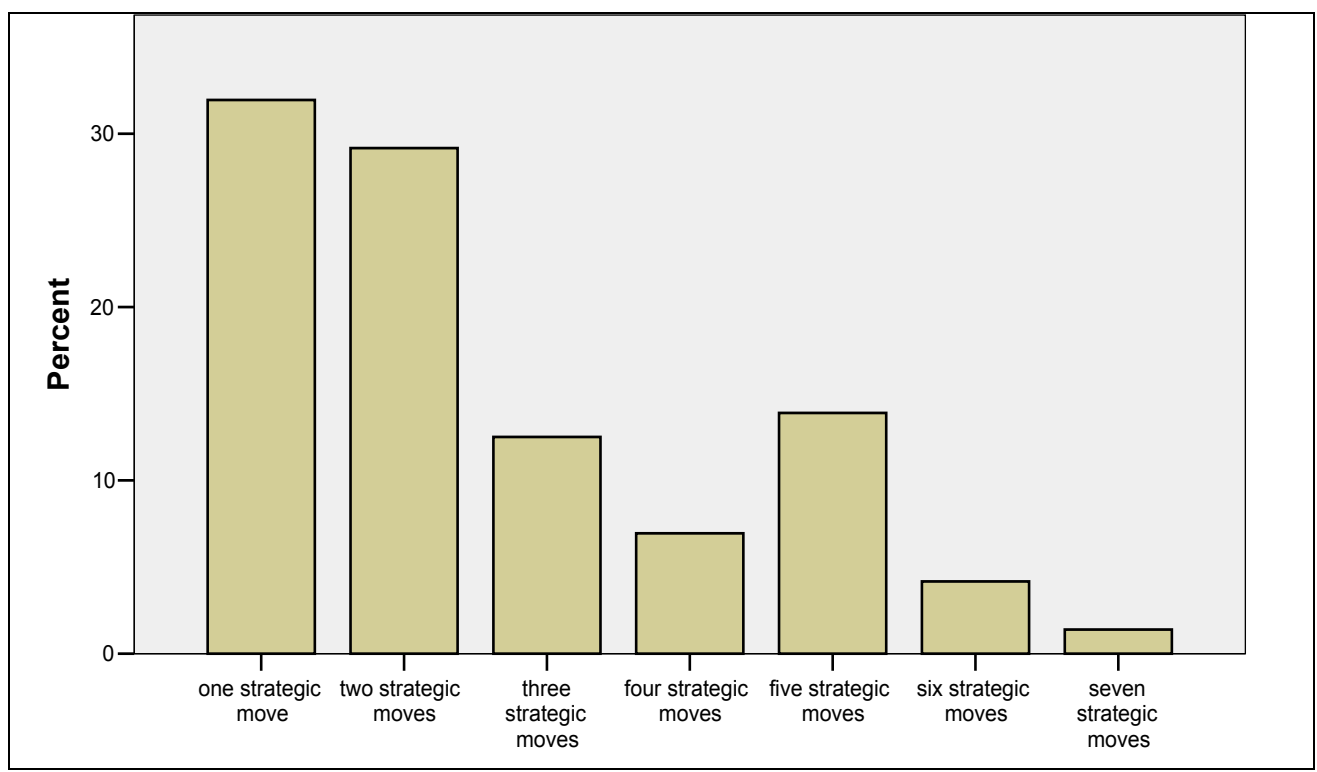

**Figure 9.** The Number of mentioned strategic moves by interviewed managers as a part of the basic bundle of overall company's development strategy

The size of the company influences the number of officially shared strategic moves by the interviewed managers. It can be detected that predominant part of big organizations has marked two, five or three strategic moves while describing the success stories of the respective entities. Since the management of small and medium sized firms is proved to be less complex than that of big organizations, it seems logical that the majority of respondents, leading SMEs share one or two strategic moves, constituting the strategic bundle of their overall company development strategies (see figure 10). Furthermore when analyzing strategic disclosure of subsidiaries, it can be stated that the managers of headquartered abroad entities shared strategic moves within the interval from four to six.

The spread of surveyed business organizations by industrial branches is shown in table 9 and figure 11. Dominating part of leading local business organizations belong to "high-tech industry", "foodstuff industry", "machine-building industry", "timbering and wood processing industry" and "transport business"

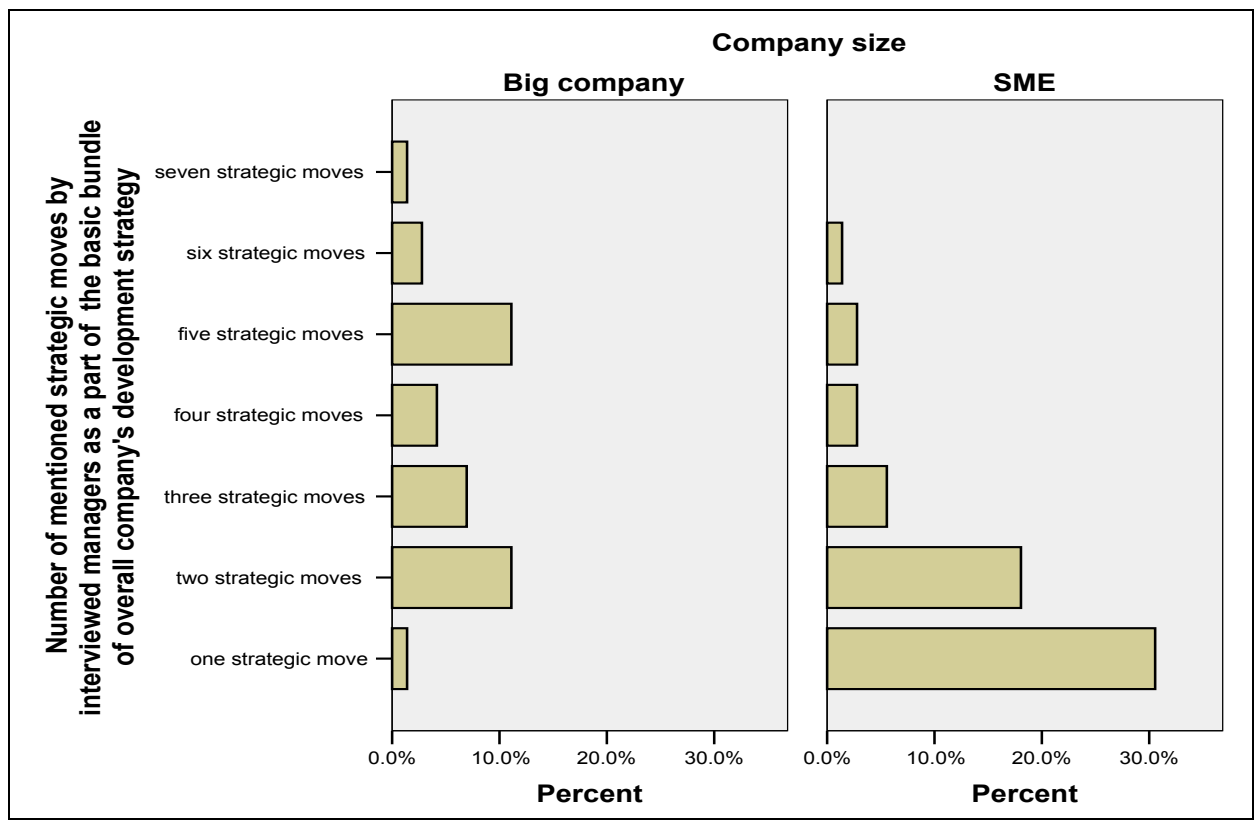

**Figure 10**. The Number of mentioned strategic moves by company's size

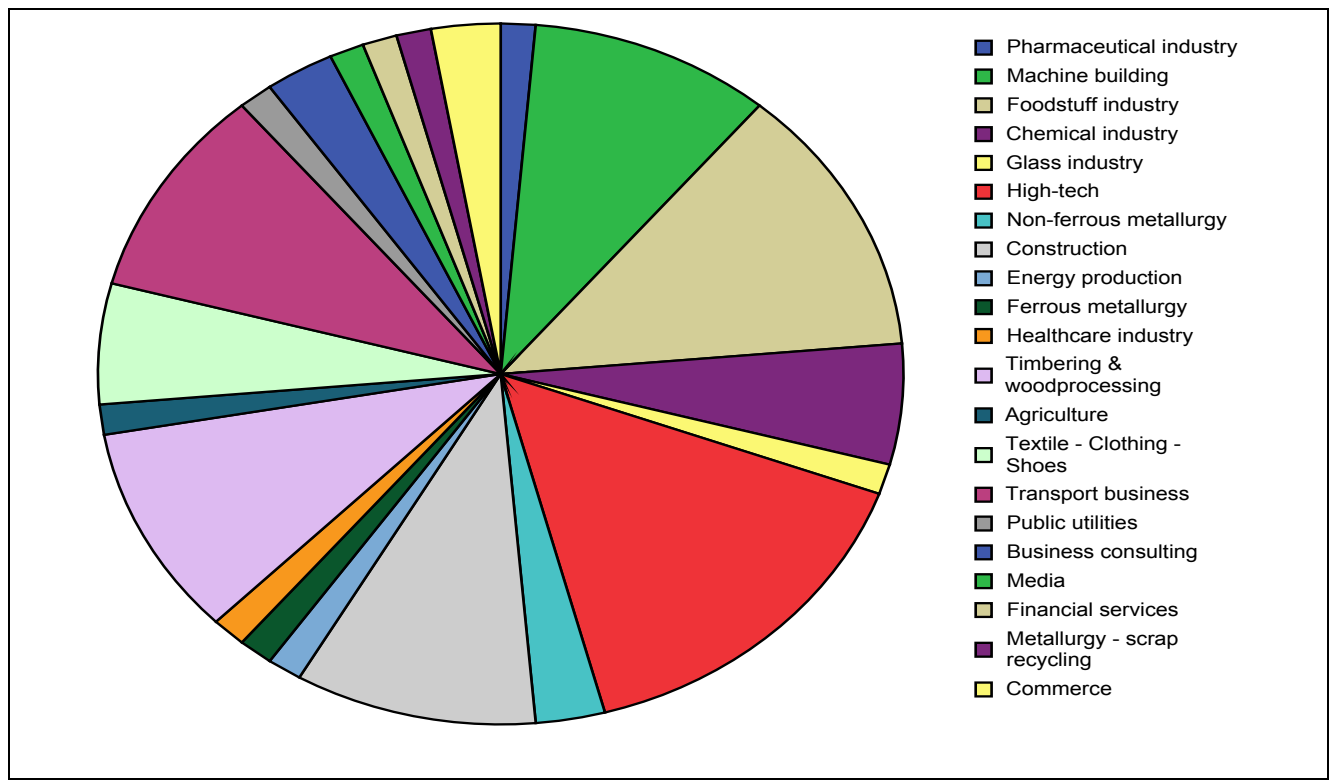

**Figure 11.** The surveyed companies by branch of industry

**Table 9.** The surveyed companies by branch of industry

|  |  | Frequency | Percent | Valid Percent | Cumulative Percent |
|---|---|---|---|---|---|
| Valid | Pharmaceutical industry | 1 | 1.4 | 1.4 | 1.4 |
|  | Machine building | 7 | 9.7 | 9.7 | 11.1 |
|  | Foodstuff industry | 9 | 12.5 | 12.5 | 23.6 |
|  | Chemical industry | 4 | 5.6 | 5.6 | 29.2 |
|  | Glass industry | 1 | 1.4 | 1.4 | 30.6 |
|  | High-tech | 11 | 15.3 | 15.3 | 45.8 |
|  | Non-ferrous metallurgy | 2 | 2.8 | 2.8 | 48.6 |
|  | Construction | 7 | 9.7 | 9.7 | 58.3 |
|  | Energy production | 1 | 1.4 | 1.4 | 59.7 |
|  | Ferrous metallurgy | 1 | 1.4 | 1.4 | 61.1 |
|  | Healthcare industry | 1 | 1.4 | 1.4 | 62.5 |
|  | Timbering & wood-processing | 7 | 9.7 | 9.7 | 72.2 |
|  | Agriculture | 1 | 1.4 | 1.4 | 73.6 |
|  | Textile - Clothing - Shoes | 4 | 5.6 | 5.6 | 79.2 |
|  | Transport business | 7 | 9.7 | 9.7 | 88.9 |
|  | Public utilities | 1 | 1.4 | 1.4 | 90.3 |
|  | Business consulting | 2 | 2.8 | 2.8 | 93.1 |
|  | Media | 1 | 1.4 | 1.4 | 94.4 |
|  | Financial services | 1 | 1.4 | 1.4 | 95.8 |
|  | Metallurgy - scrap recycling | 1 | 1.4 | 1.4 | 97.2 |
|  | Commerce | 2 | 2.8 | 2.8 | 100.0 |
|  | Total | 72 | 100.0 | 100.0 |  |

At first sight the majority of managers seem willing to proclaim further information about their organizational strategic strivings, providing unknown facts that add important characteristics to the indicated corporate or business level strategies or support the researcher's analysis (see table 10 and figure 12).

**Table 10.** Do managers mention specific facets, related with the basic bundle of overall company's development strategy?

|  |  | Frequency | Percent | Valid Percent | Cumulative Percent |
|---|---|---|---|---|---|
| Valid | No | 10 | 13.9 | 13.9 | 13.9 |
|  | Yes | 62 | 86.1 | 86.1 | 100.0 |
|  | Total | 72 | 100.0 | 100.0 |  |

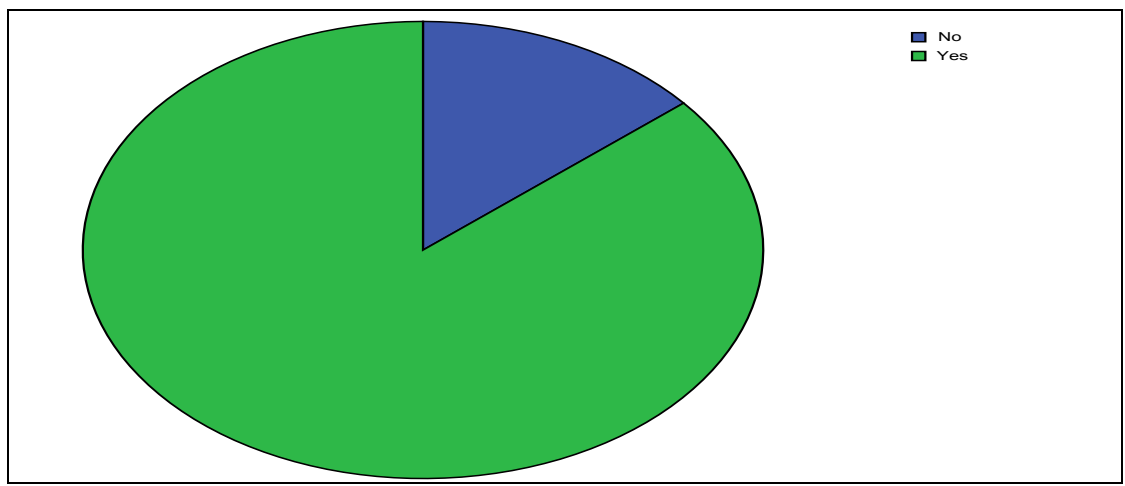

**Figure 12.** Do managers mention specific facets, related with the basic bundle of overall company's development strategy?

But a richer picture is obtained when the number of specific facets, mentioned by interviewed managers, is taken into account. In fact it showed that as a whole the respondents are not inclined to reveal publicly additional detailed facets, related with the company's strategic bundle (see table 11 and figure 13). This once again confirms the dominating cognitive paradigm in managers' minds that the area of strategic management is accepted to a great extent a traditional field of corporate secrets that should be kept for the sake of organizational survival. Furthermore, the prudent researcher cannot afford to neglect that business environment impact and always has to put up with it, while searching more creative ways of gathering needed data and applying a wider array of data retrieval and analysis methods.

**Table 11.** How many specific facets to the company's strategic bundle did the managers mention?

|  |  | Frequency | Percent | Valid Percent | Cumulative Percent |
|---|---|---|---|---|---|
| Valid | one specific facet | 29 | 40.3 | 46.8 | 46.8 |
|  | two specific facets | 17 | 23.6 | 27.4 | 74.2 |
|  | three specific facets | 11 | 15.3 | 17.7 | 91.9 |
|  | four specific facets | 5 | 6.9 | 8.1 | 100.0 |
|  | Total | 62 | 86.1 | 100.0 |  |
| Missing | System | 10 | 13.9 |  |  |
| Total |  | 72 | 100.0 |  |  |

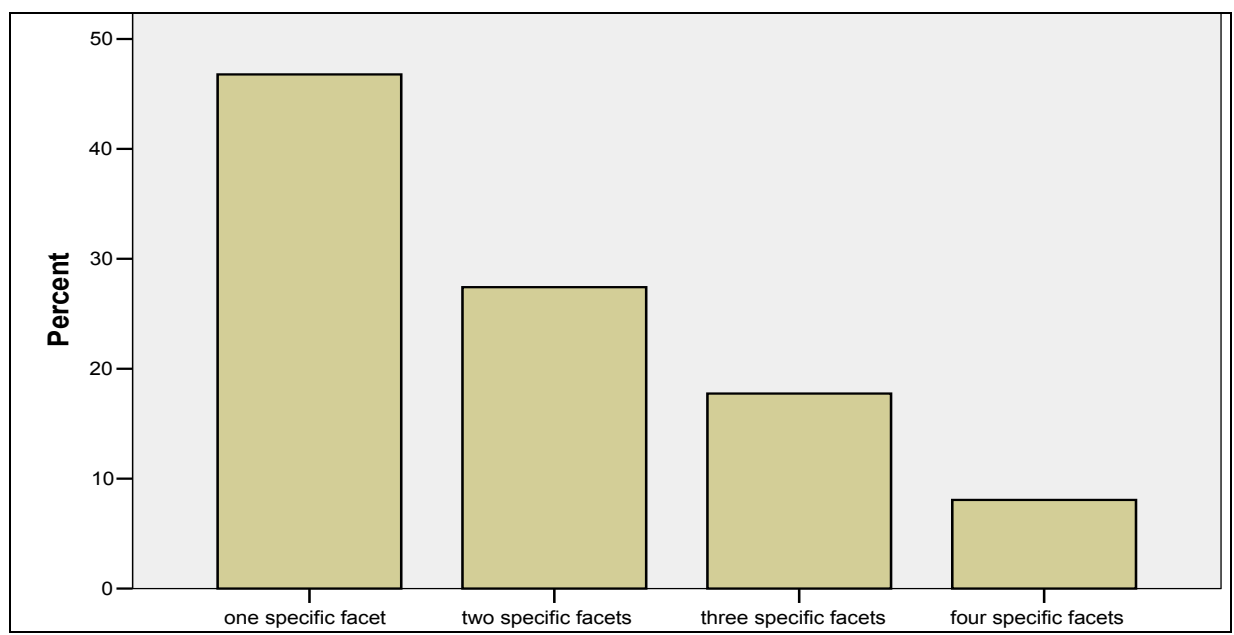

**Figure 13.** How many specific facets to the company's strategic bundle did the managers mention?

As for the applied professional language by managers it has to be marked that the content research provided evidence only for rare use of "going global", "globalization" and "turbulence". It seems that the rest of the aforementioned modern professional terms are not a part of managers' knowledge or are not shared in public.

### 5. Conclusions

A group of strategic moves, applied by succeeding business organizations in Bulgaria, may be identified. It consists predominantly by components as "Production/ service related investments", "Export", "High quality standards", "Employee training", "Innovation strategy" and "Corporate social responsibility". Some influence of three factors on the choices of strategic moves by leading organizations is detected, i.e. the availability of a mother company, its origin and entity's size. The managers of big firms shared more detailed information about their strategic moves, but as a whole "strategic management" issues still remain organizational secrets. Most of the managers do not use in public modern terms, associated with the strategic management sphere.